\documentclass[10pt,letterpaper]{article}
\usepackage{opex3}
\usepackage{subfig}
\usepackage{amsmath}
\usepackage{tikz}
\usetikzlibrary{shapes,arrows}

\begin{document}

% Define TikZ block styles
\tikzstyle{block} = [rectangle, draw, thick, text width=8.5em, text centered, rounded corners, minimum height=4em]
\tikzstyle{smblock} = [rectangle, draw, thick, text centered, rounded corners, minimum height=2em]
\tikzstyle{line} = [draw, very thick, -latex]

\title{Real-Space x-ray tomographic reconstruction of randomly
oriented objects with sparse data frames}

\author{Kartik Ayyer$^{1,*}$, Hugh T. Philipp$^1$, Mark W. Tate$^1$, Veit Elser$^1$ and Sol M. Gruner$^{1,2,3}$}

\address{$^1$Cornell University, Laboratory of Atomic and Solid State Physics,\\
Ithaca, NY, USA\\
$^2$Cornell's High Energy Synchrotron Source (CHESS), Ithaca, NY USA\\
$^3$Kavli Institute at Cornell for Nanoscale Science, Ithaca, NY, USA}

\email{$^*$ksa47@cornell.edu}

\begin{abstract}
Schemes for X-ray imaging single protein molecules using new x-ray sources, like x-ray free electron lasers (XFELs), require processing many frames of data that are obtained by taking temporally short snapshots of identical molecules, each with a random and unknown orientation. Due to the small size of the molecules and short exposure times, average signal levels of much less than 1 photon/pixel/frame are expected, much too low to be processed using standard methods. One approach to process the data is to use statistical methods developed in the EMC algorithm (Loh \& Elser, Phys. Rev. E, 2009) which processes the data set as a whole. In this paper we apply this method to a real-space tomographic reconstruction using sparse frames of data (below $10^{-2}$ photons/pixel/frame) obtained by  performing x-ray transmission measurements of a low-contrast, randomly-oriented object. This extends the work by Philipp et al. (Optics Express, 2012) to three dimensions and is one step closer to the single molecule reconstruction problem.
\end{abstract}


\section{Introduction}

X-ray free electron lasers (XFELs) have been successful in performing crystallographic reconstructions of biomolecules with nanocrystalline samples having as few as $10^3$ unit cells\cite{chapman2011}\cite{boutet2012}. This is possible, in part, because the high signal and Bragg peak concentration allows indexing methods to determine the orientation of each frame \cite{white2012}.  There are, however, many situations in which indexing with a single frame is not possible either because of the nature of the sample (e.g. a non-crystalline particle or protein) or because the number of scattered photons detected in a single frame simply do not provide enough information.  In these cases, a different approach is required.

The EMC (expand-maximize-compress) algorithm\cite{loh026705}\cite{loh2010} presents a method of dealing with these more difficult data sets.  The heart of this algorithm depends on expectation maximization (EM) methods that were experimentally demonstrated previously\cite{philipp2012} using the prototype detector based upon the ASIC (application specific integrated circuit) developed by Cornell for the CS-PAD detector at the Linac Coherent Light Source (LCLS). 

The present work extends this approach to 3D tomographic reconstructions using sparse x-ray transmission data collected from a 5 cm sized object, where each data frame is from a random and unknown orientation. The sparse data frames used for reconstruction have signal levels of ${\sim 10^{-3} - 10^{-2}}$ photons/pixel/frame.   The detector system used is a tiled pixel array detector (PAD).  Each tile has 128$\times$128 pixels and uses the mixed-mode pixel array detector (MMPAD) ASIC\cite{vernon2007}.  The overall tiling format is a 3$\times$2 grid\cite{tate2013}. The object reconstructed is relatively low-contrast at the $K_{\alpha}$ emission line of molybdenum (17.5 keV). The object absorbs from 5\% of the photons
incident on it in the thinnest regions to 90\% in the central portion.

\section{Data generation}
\label{sec:datagen}

\begin{figure}
\centering
\subfloat[]{
    \label{fig:ninjago}
    \includegraphics[width=\textwidth]{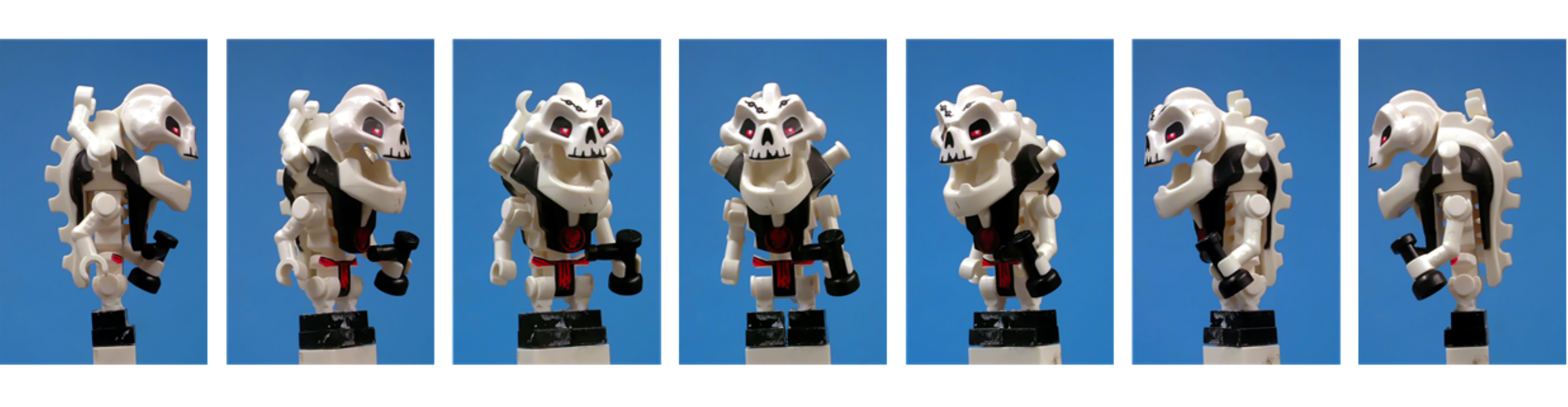}
}\\
\subfloat[]{
    \label{fig:frame1}
    \fbox{\includegraphics[width=0.3 \textwidth]{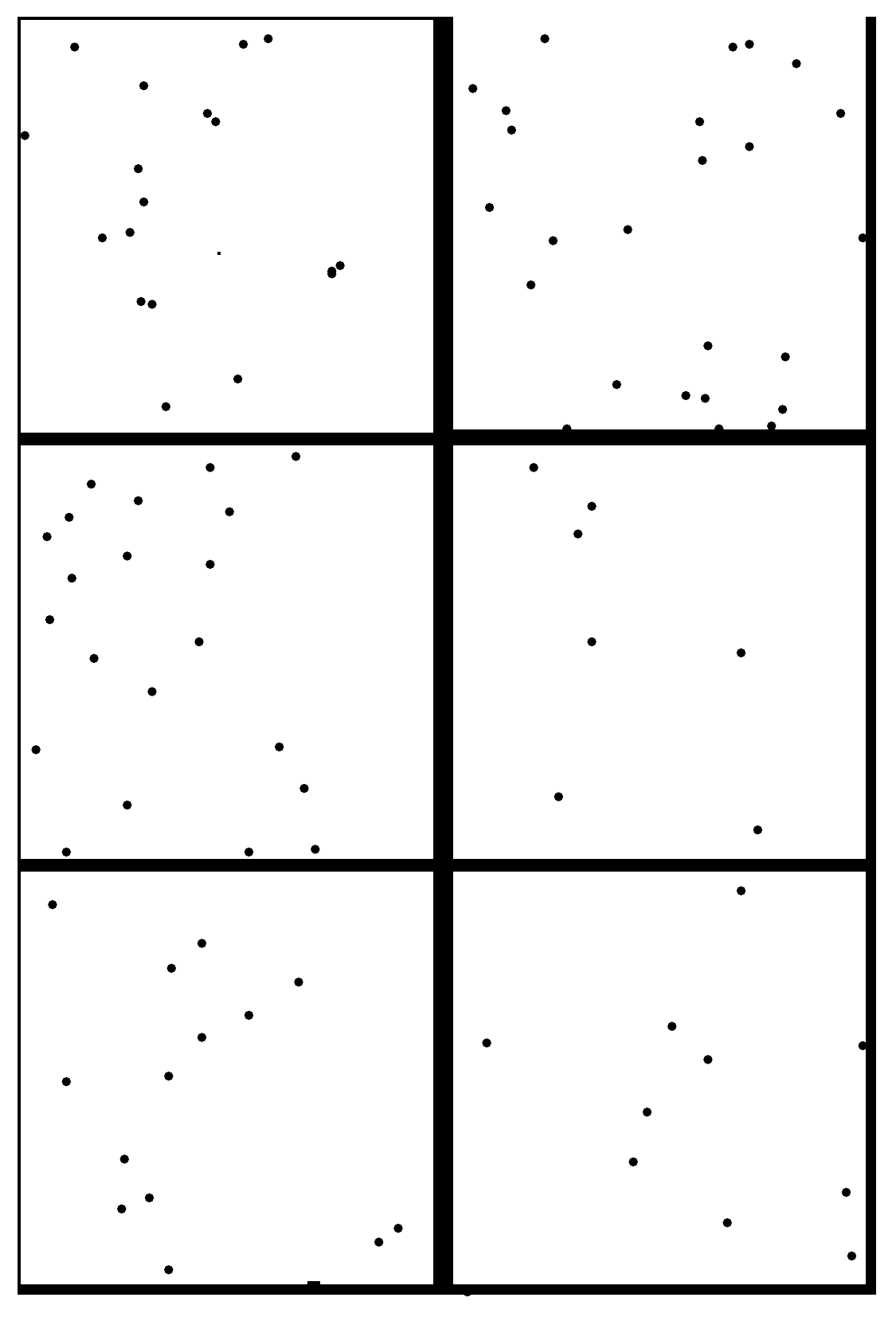}}
}
\subfloat[]{
    \label{fig:frame10}
    \fbox{\includegraphics[width=0.3 \textwidth]{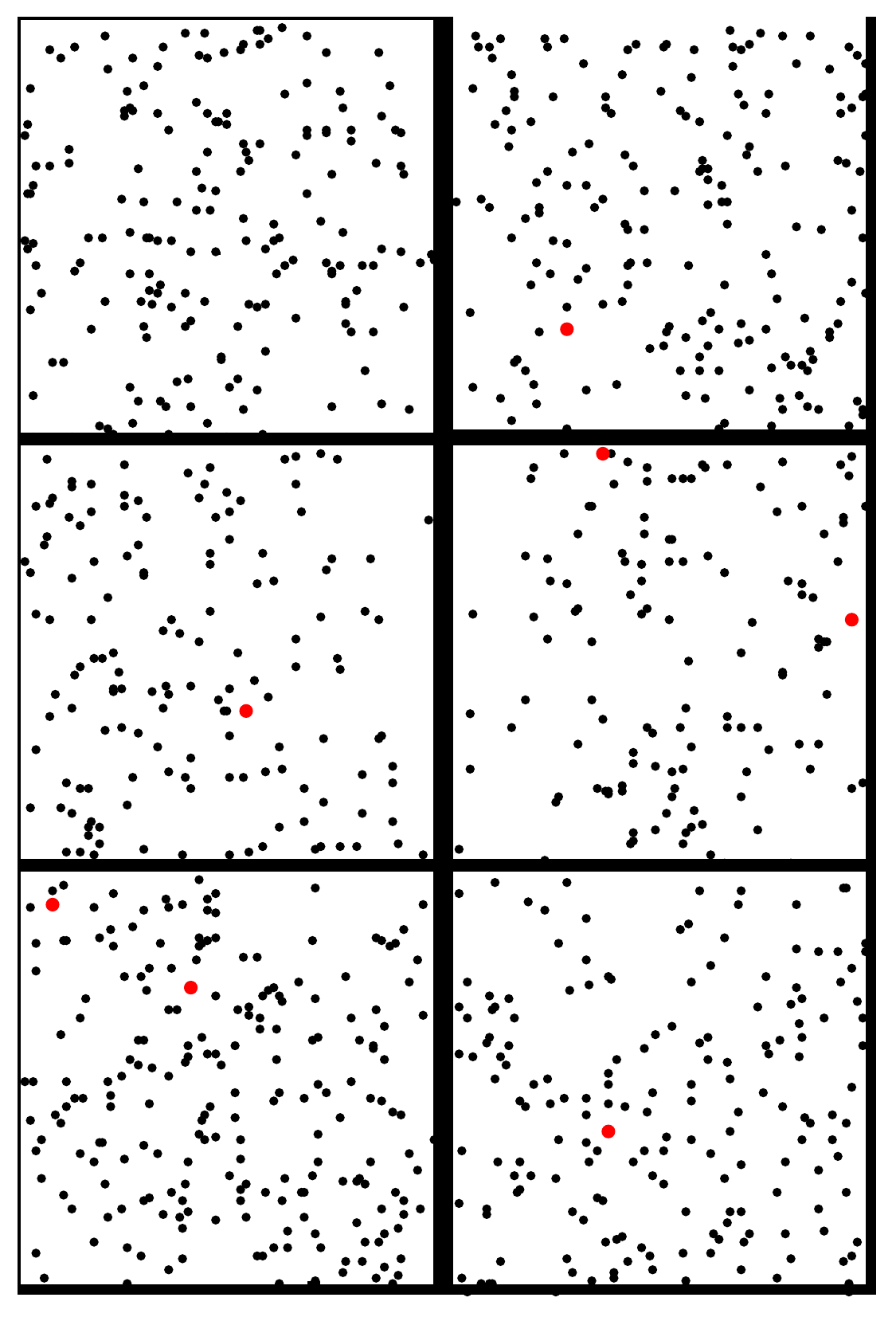}}
}
\caption{(a) Photographs of the target object, a Lego NINJAGO Samukai\textregistered~figure. (b) Typical frame of data containing 96 photons in a 396x266 pixel detector. This translates to \mbox{$9.1\times10^{-4}$} photons/pixel. (c) Data frame with 1025 photons obtained by combining 10 consecutive frames from the previous data set. The sizes of the pixels recording photons have been enlarged to improve visibility. Pixels with two photons are shown in red.}
\end{figure}

Low-signal tomographic reconstructions using the EMC algorithm were tested using a plastic Lego NINJAGO Samukai\textregistered~figure, roughly matching the size of the detector, as the object of study. A 50 watt molybdenum anode x-ray tube was used to generate x-rays (Tru-Focus TCM-5000M). The applied high voltage was 21.5 kV and the current was set to 0.05 mA. A 400 micron zirconium filter was used to better isolate the Kα line of molybdenum (17.5 keV) from the x-ray tube emission spectrum by preferentially attenuating lower energies and higher energies (beyond the K-edge of zirconium, 18.0 keV). The distance between the x-ray source and the sample was 1.3 meters. The detector was placed directly after the sample. Detector pixels measured 150 $\mu$m $\times$ 150 $\mu$m. The per-frame exposure time was chosen to be 4 ms with an average signal of 100 photons per frame ($10^{-3}$ photons/pixel/frame).

The NINJAGO figure (Figure~\ref{fig:ninjago}) was mounted on a post and attached to a rotation stage (Newport URS100BPP).  During data taking, the rotation stage turned continuously at a rate of 2 degrees per second.  The data was taken continuously with 860 $\mu$s between frames in batches of 1000 frames.  Variable time delays were used between batches, to ensure that there was no time sequence bias, and 15.6 million frames were acquired.

Each frame (Figure~\ref{fig:frame1}) acquired was converted to photon counts in the following way: 1) An average of 100 to 300 dark frames was subtracted from the signal frames and detector specific offsets were corrected. 2) A threshold was applied to each pixel that corresponded to 60\% of a single molybdenum K$_\alpha$ x-ray. Pixels that did not exceed this threshold were set to zero signal level. 3) The number of photons detected in a pixel exceeding the threshold was determined by dividing the pixel signal by the single photon signal level, and rounding to the nearest integer. Note the edge pixels around the rim of each tile are larger than the interior pixels due to edge effects in the sensor\cite{green2013}. No correction for this effect was applied to the data.

After digitizing each frame, a list of pixels having non-zero photons was recorded.  Because of the sparse nature of the low-fluence data, recording pixel coordinate and number of photons, rather than all pixel values, greatly reduced the memory required to store data. These data frames were passed to the algorithm without any information about the rotation of the object corresponding to each frame. This was done to simulate the unknown-orientation aspect of the single molecule imaging process.

Three more data sets were generated by combining 2, 4, and 10 consecutive frames within the 1000 frame batches, respectively. Since the object rotated approximately $8\times10^{-3}$ degrees between frames, we can safely assume that 10 consecutive frames are from essentially identical orientations of the object.

\section{Pre-processing of data}
\label{sec:preprocess}

\begin{figure}
    \centering
    \subfloat[]{
        \label{fig:powder}
        \includegraphics[width=0.4\textwidth]{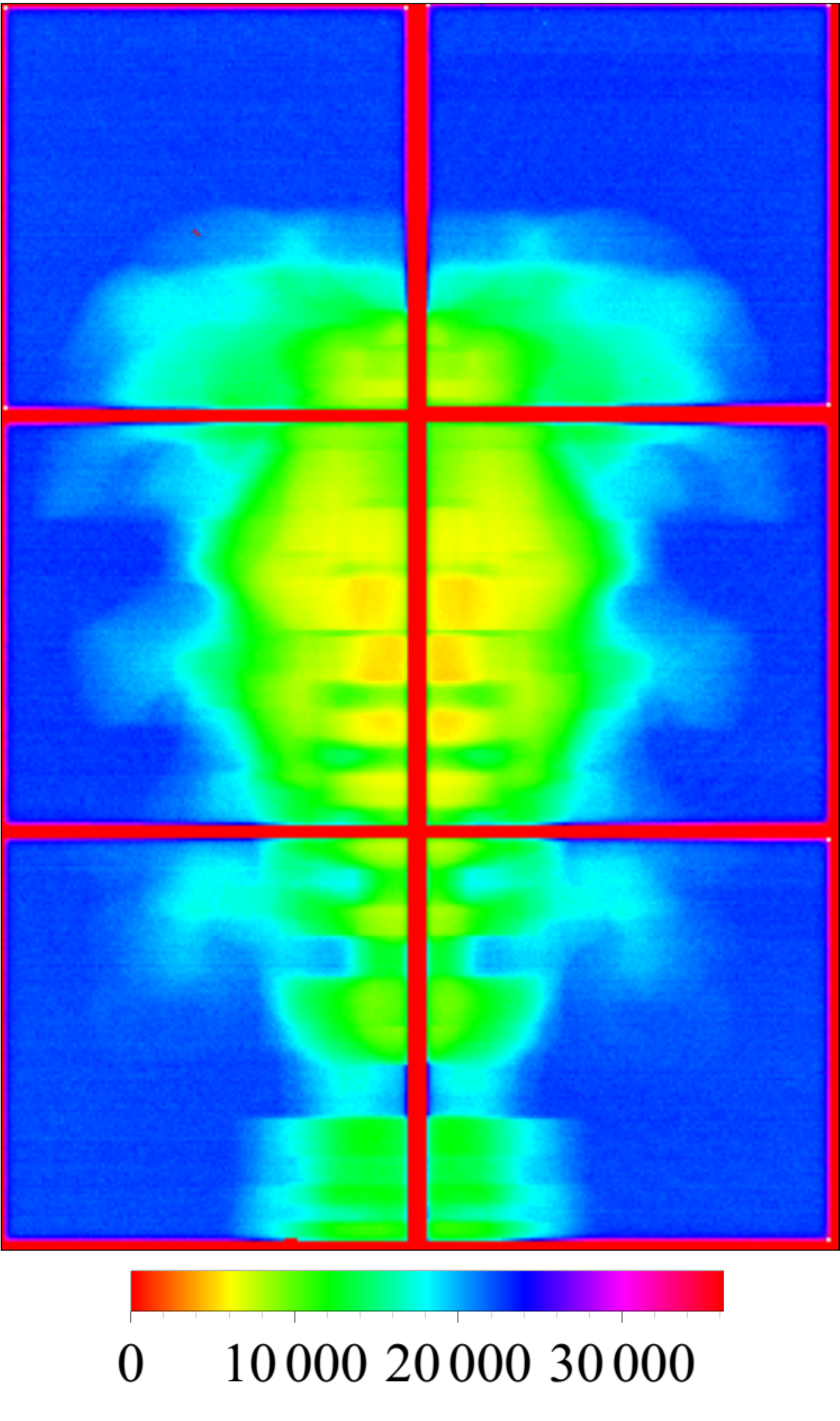}
    }
    \subfloat[]{
        \label{fig:mask}
        \includegraphics[width=0.4\textwidth]{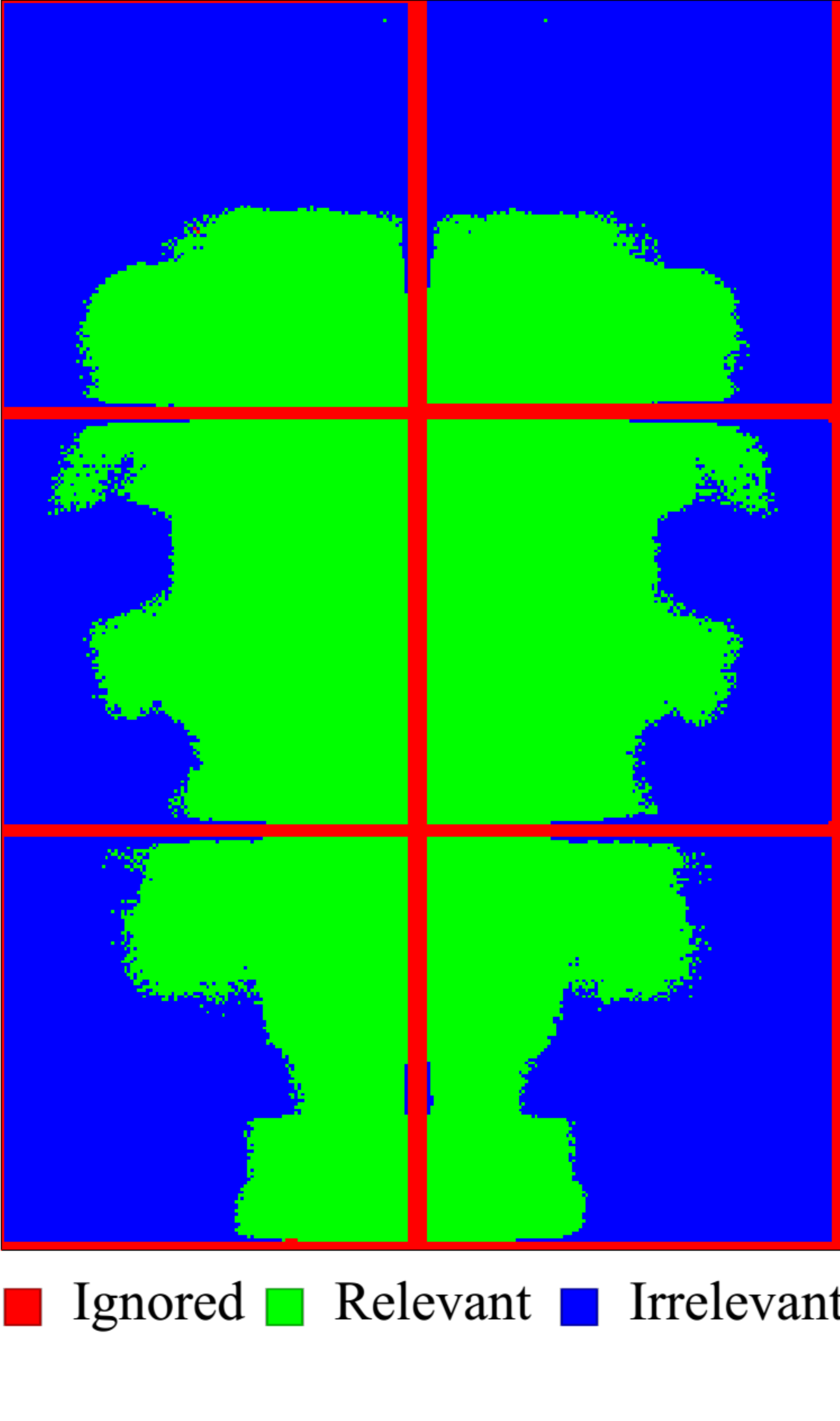}
    }\\
    \subfloat[]{
        \centering
        \label{fig:histogram}
        \includegraphics[width=0.8\textwidth]{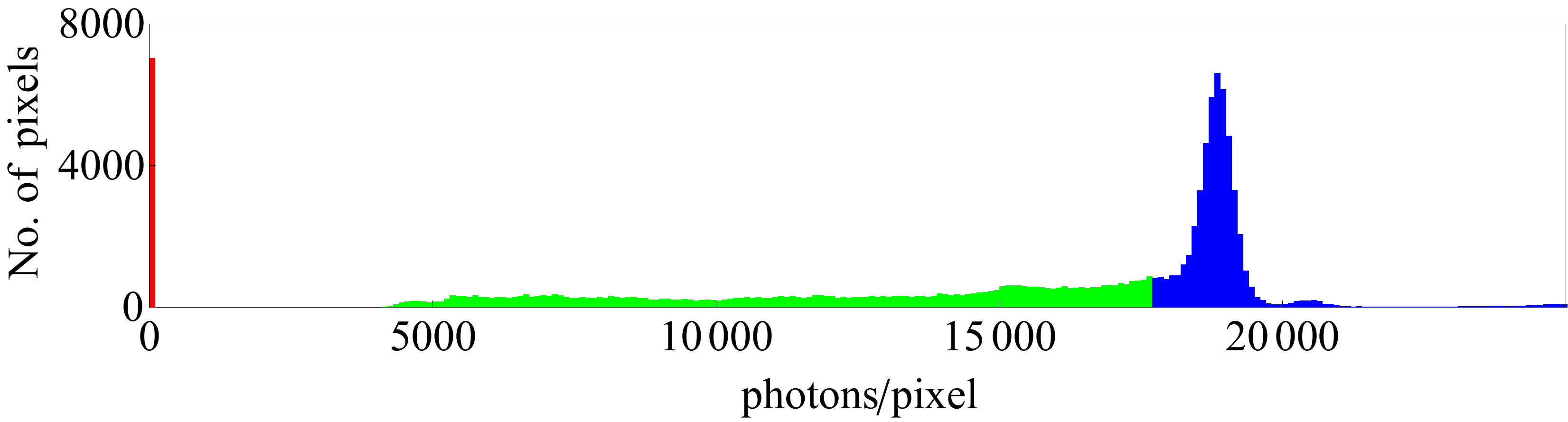}
    }
    \caption{(a) Angle-averaged pattern produced by summing over all 15,650,615 frames in the 99 photons/frame data set. The numbers in the legend refer to photon counts. (b) Mask representing relevant (green), irrelevant (blue) and ignored (red) pixels (details in Section~\ref{sec:preprocess}) (c) Histogram of photon counts with a cutoff value for relevant pixels at 17,800 photons.}
    \label{fig:preprocess}
\end{figure}

Figure~\ref{fig:powder} shows the angle-averaged pattern obtained by summing over all frames. Since pixels in the outer region (e.g. top-right and top-left corner) are never obscured by the object, they provide no structural or orientational information. From the histogram in Fig.~\ref{fig:histogram}, a photon count of 17,800 was chosen as the cutoff to define a mask of pixels as shown in Fig.~\ref{fig:mask}. Only the green pixels are used to determine the orientations of each frame of data. In the remainder of the paper, these pixels are referred to as relevant and the others as irrelevant.

Another mask (red in Fig.~\ref{fig:mask}) is used to exclude the pixels in the gaps between detector tiles. If this is not done, the algorithm naturally interprets these gaps as coming from an infinitely opaque structure obstructing the view in every frame. Since there are no photons here, the exact attenuation caused by this structure is undefined except that it is above a certain level. In addition to these pixels, this mask also includes 7 ``hot" pixels that were malfunctioning and erroneously record extremely high count rates. This mask of ignored pixels is shown in red in Fig.~\ref{fig:mask}. The pattern is not symmetric or smooth because of statistical noise, small variations in pixel gain, detection efficiency and small errors in pixel offsets.  The lack of symmetry and smoothness present no problem for data reduction and the algorithm is generally robust in this regard. 

\section{Reconstruction algorithm}

\begin{figure}[ht]
    \centering
    \begin{tikzpicture}[node distance = 4.5cm, auto]
        \node [block] (expand) {Expand:\\$\widetilde{M}_{uvw} \rightarrow W_{rz} (\theta)$};
        \node [block, right of=expand] (maximize) {Maximize:\\$W_{rz}(\theta)  \rightarrow W'_{rz}(\theta)$};
        \node [block, right of=maximize] (compress) {Compress:\\$W'_{rz}(\theta) \rightarrow \widetilde{M}'_{uvw}$};
        \path [line] (expand) -- (maximize);
        \path [line] (maximize) -- (compress);
        \draw [line] (compress) to[out=-120,in=-60] (expand);
    \end{tikzpicture}
    \caption{Flowchart of EMC reconstruction algorithm applied to this system including the transformations performed in each step.}
    \label{fig:emc}
\end{figure}
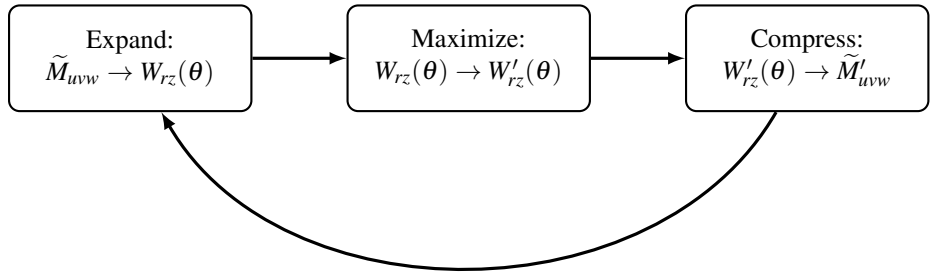

The imaging process can be thought of as unknown-angle tomography at very low signal. This problem naturally splits into two parts, tomography and determination of angle. The projection-slice theorem was used to tackle the first and an iterative expectation-maximization (EM) based algorithm for the second.

Since the projection-slice operation is applied partly in Fourier space, it is convenient to have the iterate be a 3D Fourier space model $\widetilde{M}_{uvw}$. The 3D inverse Fourier transform of $\widetilde{M}_{uvw}$

\begin{equation}
\mathrm{3DIFT}[\widetilde{M}_{uvw}]=M_{xyz}
\end{equation}
gives the attenuation, by $\exp(-M_{xyz})$, of x-rays passing through voxel $xyz$ of the object. 

The relevant pixel mask is used to generate the initial random model. Consider the three-dimensional object generated by rotating this mask about the axis of rotation. Since we assume that all pixels outside the relevant region are never obscured for any angle, this `rotated-mask' object acts as the support for the target object. Thus, voxels inside this object are assigned a random number uniformly in the range [0,1] and the voxels outside are zeroed. This 3D array is zero-padded perpendicular to the rotation axis to reduce interpolation errors and Fourier transformed to generate the initial random Fourier model.

We express our algorithm in the Expand-Maximize-Compress (EMC) framework of \cite{loh026705}. Starting with the initial random model, in each iteration, these three operations are applied to it to generate the updated model. Here the Expand and Compress steps represent transformations between the 3D attenuation model in Fourier space and the collection of 2D real-space intensity attenuation patterns, for uniformly sampled object-rotation angles. We will refer to the latter as tomograms. The Maximize step generates updated tomograms which increase the likelihood of the data being generated from the model. Thus, in each iteration, we generate the tomograms from the current model (Expand), update the tomograms (Maximize), and combine them to generate the new model (Compress).

\subsection{Expand}
\label{sec:expand}

\begin{figure}[ht]
    \centering
    \begin{tikzpicture}[node distance=4cm,auto]
        \node [smblock] (muvw) {$\widetilde{M}_{uvw}$};
        \node [smblock, right of=muvw] (trhow) {$\widetilde{T}_{\rho w}(\theta)$};
        \node [smblock, right of=trhow] (trz) {$T_{rz}(\theta)$};
        \node [smblock, right of=trz] (wrz) {$W_{rz}(\theta)$};
        \path [line] (muvw) -- node{Interpolation} (trhow);
        \path [line] (trhow) -- node{2DIFT} (trz);
        \path [line] (trz) -- node{Exponentiation} (wrz);
    \end{tikzpicture}
    \caption{The Expand step generates tomograms, $W_{rz}(\theta)$, for many different discrete orientations $\theta$ from the 3D Fourier space model $\widetilde{M}_{uvw}$.}
    \label{fig:expand}
\end{figure}
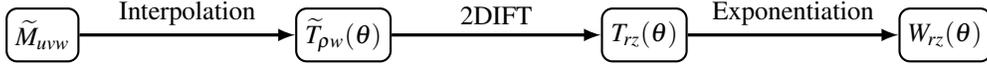

The tomograms are generated using the projection-slice theorem augmented by exponentiation of the projections. We use linear interpolation to generate slices $\widetilde{T}_{\rho w}(\theta)$ of $\widetilde{M}_{uvw}$ passing through the axis $(u,v) = (0,0)$ for a large number of uniformly spaced orientations, $\theta$. To avoid interpolation errors, we oversample the Fourier space model. Thus, if there are $U\times W$ pixels in the detector and the rotation axis is along the $w$-axis, the iterate has $(\sigma U) \times(\sigma U)\times W$ complex voxels, where $\sigma$ is the oversampling factor.

We inverse Fourier transform the slices $\widetilde{T}_{\rho w}(\theta)$ to generate the projected attenuations $T_{rz}(\theta)$ and then generate the intensity models by applying the formula,

\begin{align}
    T_{rz}(\theta) &= \mathrm{2DIFT}[\widetilde{T}_{\rho w}(\theta)] \\
    W_{rz}(\theta) &= f . \exp[-s T_{rz}(\theta)]
    \label{eqn:exp}
\end{align}
Here, $f$ represents the unattenuated intensity and $s$ represents a scale factor explained below.. Since we assume that the irrelevant pixels are unobstructed in any orientation, we can obtain $f$ from the mean photon count in these pixels. 

Before we apply the Maximize step on $W_{rz}(\theta)$, we must determine if we have the right overall scale for $T_{rz}(\theta)$. From the data, we know the mean number of photons/frame, $\langle \sum_{rz}W_{rz}(\theta)\rangle$, where the angle bracket denotes averaging over all orientations $\theta$. However, due to the exponentiation in Eq.~(\ref{eqn:exp}), we cannot use this to directly calculate $\langle T_{rz}(\theta)\rangle$. We also observe that the algorithm is prone to a scaling instability in the first few iterations where the values in $T_{rz}(\theta)$ explode, leading to underflows in the intensity due to Eqn~\ref{eqn:exp}. This is avoided by numerically determining the correct scale every iteration. We use the secant method~\cite{weissteinsecant} to solve

\begin{equation}
0 = \langle W_{rz}(\theta)\rangle - f . \langle \exp[-s T_{rz}(\theta)]\rangle
\end{equation}
for the unknown scale factor $s$.

\subsection{Maximize}

In this step we find the updated intensities $W'_{rz}(\theta)$ using expectation maximization. Due to the low signal count per pixel, we expect the probability of a photon incident on a pixel to be governed by Poisson statistics. At each pixel, the intensity model value gives the mean of this Poisson distribution. Thus, the likelihood of a frame of data $d$ ($K_d$) coming from $W_{rz}(\theta)$ is given by,

\begin{align}
    \ell_d(\theta) &= \prod_{rz} \frac{W_{rz}(\theta)^{K_{d,rz}} \exp[-W_{rz}(\theta)]}{K_{d,rz}!}
    \label{eqn:like}\\
    \Rightarrow p_d(\theta) &= \frac{\ell_d(\theta)}{\sum\limits_\theta \ell_d(\theta)} 
    \label{eqn:prob}
\end{align}
where $p_d(\theta)$ is the probability obtained by normalization and $K_{d,rz}$ is the number of photons at pixel $(r,z)$ in frame $d$. The $K_{d,rz}!$ factor in the denominator of Eq.~\ref{eqn:like} cancels out in the calculation of $p_d(\theta)$. To increase the effectiveness of our probability assignment, we only consider photons in relevant pixels. 

Using these probabilities we calculate the updated intensities, $W'_{rz}(\theta)$, which maximize the log-likelihood of generating the data,
\begin{equation}
\log[Q(W'_{rz}(\theta))] = \sum_d \sum_\theta \sum_{rz} p_d(\theta) [K_{d,rz} log(W'_{rz}) - W'_{rz}]
\end{equation}
Rearranging the sums and maximizing with respect to $W'_{rz}$, we get
\begin{equation}
W'_{rz}(\theta) = \frac{\sum\limits_d p_d(\theta) K_{d,rz}}{\sum\limits_d p_d(\theta)}
\end{equation}

For two reasons explained below, we apply an ``inertia" factor $\alpha$ in the update rule for $W_{rz}(\theta)$. We apply an update rule, 
\begin{equation}
W'_{rz}(\theta) \leftarrow \alpha W_{rz}(\theta) + (1 - \alpha) W'_{rz}(\theta)
\end{equation}
Thus, $W'_{rz}(\theta)$ has a contribution from the previous iteration.

In the first few iterations, where the iterate changes rapidly, a high value of $\alpha$ prevents $T_{rz}(\theta)$ from exploding, leading to underflows in the intensity calculation. Secondly, it also provides an additional handle on the rate of convergence of the algorithm. In the current geometry, the various slices $\widetilde{T}_{\rho w}(\theta)$ do not overlap each other except at the rotation axis. This means that there is only a weak constraint for successive slices to correspond to successive rotation angles. Thus, there are near solutions which have arbitrary jumps in angles in successive slices. Indeed, this is what we observe we converge to when we have relatively high signal. However, if we slow down the rate of convergence using a high inertia factor, we reliably converge to the right solution. We do not expect this to be an issue if we have full 3D rotation as the various slices strongly overlap. In that case, the Expand and Compress steps should together provide a strong constraint which impose the correct ordering on the slices.

\subsection{Compress}

This is the inverse of the Expand step. First, we generate the attenuation projections $T'_{rz}(\theta)$ by taking the negative logarithm of the updated tomograms,

\begin{equation}
T'_{rz}(\theta) = -\log(W'_{rz}(\theta)/f))
\end{equation}
The ignored pixels from the panel gaps and ``hot" pixels are not updated. These updated projections are then zero-padded and Fourier transformed to get the updated slices $\widetilde{T'}_{\rho w}(\theta)$. We then use linear interpolation to generate the updated 3D model.

\section{Results}

\begin{table}
    \centering

    \begin{tabular}{| c | c | c | c | c |}
        \hline
        & \parbox[t]{2cm}{\centering Total\\ photons/frame} & \parbox[t]{2cm}{\centering No. of frames\\ (millions)} & \parbox[t]{2cm}{\centering Relevant\\ photons/frame} & \parbox[t]{2cm}{\centering Absorbed\\ photons/frame}\\
        \hline
        1 & 99.5 & 15.6 & 37.0 & 3.9\\
        2 & 198.7 & 7.8 & 73.8 & 7.8\\
        3 & 397.2 & 3.9 & 147.6 & 15.6\\
        4 & 991.6 & 1.6 & 368.4 & 39.1\\
        \hline
    \end{tabular}
    \caption{Parameters of four data sets analyzed. The last three were generated by combining 2, 4, and 10 successive frames of the first data set. Relevant pixels refer to the green region in Fig.~\ref{fig:mask}. Only the photons in this region have orientational information about the data frames.}
    \label{tab:data}
\end{table}

Data was taken with 99.5 mean photons per frame. Of these, only 37 photons were incident in the relevant region of the detector as defined in Section~\ref{sec:preprocess}. Using the mean incident flux calculated from the irrelevant pixels, we can determine that on average around 4 photons were absorbed by the object per frame. As mentioned in Section~\ref{sec:datagen}, frames were grouped by either 1, 2, 4, or 10 consecutive frames into 4 data sets. After this initial grouping, no information on the angular position of the combined frames was passed to the reconstruction algorithm. Table~\ref{tab:data} lists the properties of all four data sets.

\begin{figure}
    \centering
    \includegraphics[width = \textwidth]{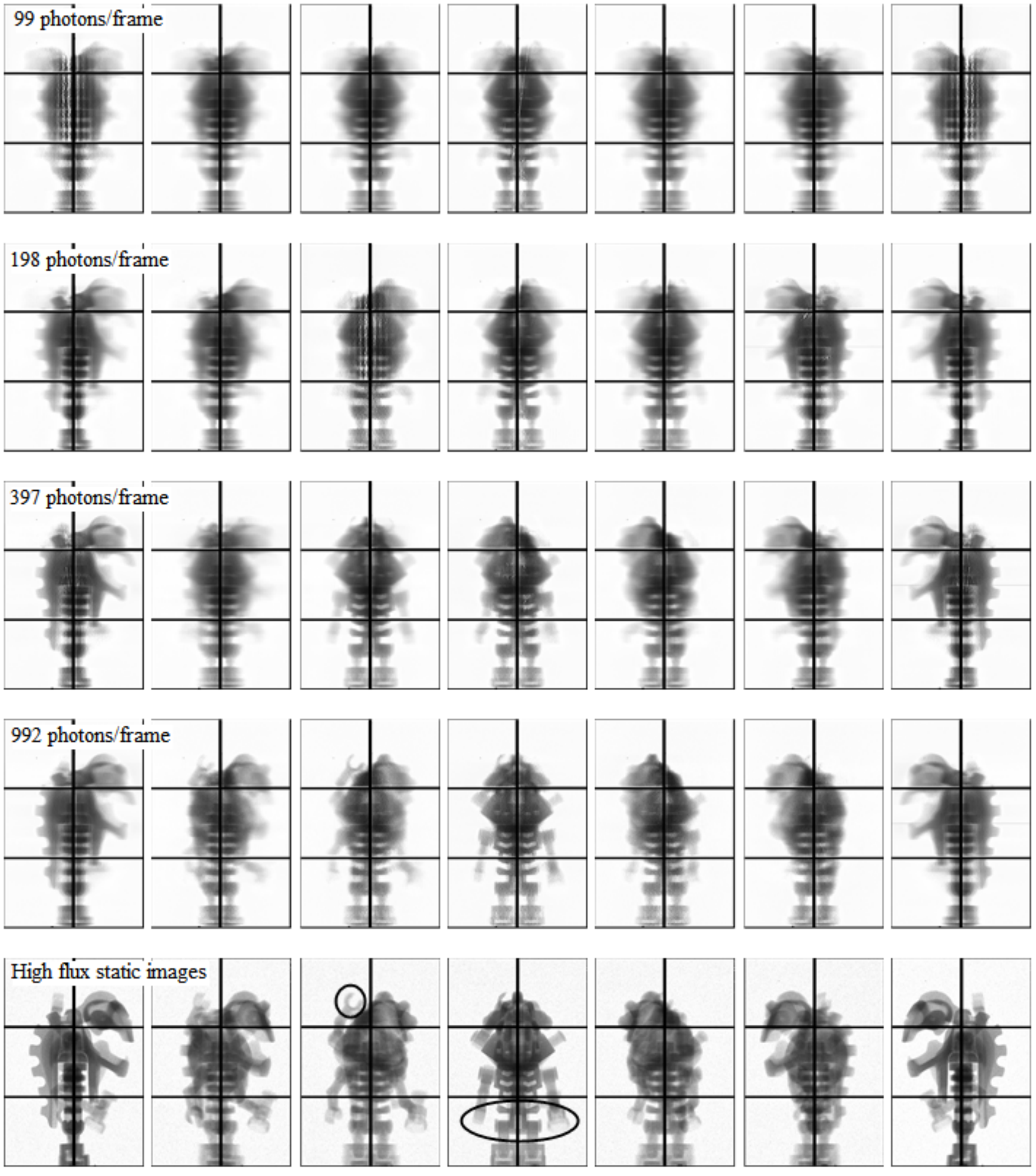}
    \caption{The first four rows show the projected x-ray transmission intensities through the reconstructed object for $\pi/6$ rotation intervals in $[0,\;\pi]$. Reconstructions of the object were obtained with data sets of 99, 198, 397, and 992 photons per frame, respectively. The total number of photons in each data set was $1.56\times10^9$. Details about the data sets are given in Table~\ref{tab:data}. The bottom row shows for comparison static radiographs of the object which were acquired at the same angles at high signal levels. Some fine, low-contrast features are circled. All images are scaled such that white and black colors represent no and complete attenuation respectively.}
    \label{fig:results}
\end{figure}

Figure~\ref{fig:results} shows reconstructions from the four data sets along with a set of high flux, static projections. In all four cases, the only thing that changes is the signal per frame. The total number of photons and the object is unchanged. We can clearly see that the quality of the reconstruction improves as we increase the mean number of photons/frame. With 99 photons/frame, the algorithm is unable to determine even the gross shape of the object. The finer, low-contrast features are reconstructed with more and more accuracy as we increase the number of frames combined. In the bottom row, the circle shows the location of the extra upper arm, which was attached only on one side. The oval shows the asymmetry in the two lower arms due to one of them holding a Lego dumbbell.

Note that even with 10 frames combined, there are still only $7.9\times10^{-3}$ photons/frame/pixel in the relevant region of the detector.

\section{Methods}

For all four data sets, an oversampling factor, $\sigma$, of 2 was chosen. Thus, the Fourier-space iterate had dimensions $396\times532\times532$.The angular range was divided into 200 discrete orientations and the inertia factor was chosen to be 0.8 in each case. A cutoff count of 17,800 was used to generate the relevant pixels mask. There were 46,890 relevant pixels and 8,670 ignored pixels.

Due to the large size of the problem, a parallel algorithm was required. The most time-consuming step in each iteration was implementing Equation~\ref{eqn:like}. The parallelization was applied over the $\theta$ index i.e. each process was assigned a part of the angular range for which to calculate $p_d(\theta)$. Reconstructions were performed at the LCLS cluster at SLAC. With 120 processes on 10 nodes, an iteration took 2-3 minutes and the algorithm took 30--50 iterations to converge. To test for convergence, reconstructions were performed multiple times with different random starts and they yielded the same result.

\section{Conclusions}

We view these results as an extension of the 2D reconstructions performed previously with similarly sparse data in \cite{philipp2012}. The in-plane rotation axis in the present study meant that we had to reconstruct a 3D object. Any given frame in this experiment did not have the full structural information. Due to the large size of the object, a non-linear attenuation model had to be used. Also, we needed a parallel code running on a cluster to perform the reconstruction in a reasonable time. With this experiment, we are one step closer to demonstrating the reconstruction of the 3D intensity of a biomolecule or nanocrystal in conditions of very low signal.

There are further avenues that could be pursued to improve the quality of the reconstruction: 1) One could take into account the small, but finite divergence of the beam. This would make the analysis in the expand and compress steps similar to fan-beam tomography. 2) The data at the edge pixels could be modified to reflect their larger size. 3) The axis of rotation was assumed to be aligned along the middle of the detector. This could estimated more accurately. 4) Our criterion for the relevant pixel mask was chosen for simplicity (a hard cutoff on the number of photons/pixel). This could be further refined to maximize the capacity to determine the orientation of a frame.

In the process of constructing a 3D proof-of-principle experiment for biomolecule imaging with x-rays, our experiment closely approaches the setup of cryo-electron microscopy (cryo-EM) experiments\cite{scheres2007} for biomolecules. In that case, we also have tomography with unknown orientations at very low signal-to-noise. However, there are a few differences. The cryo-EM reconstructions do not have the assistance of a rotation axis, and so have molecules in all orientations in SO(3). In addition, they must also fix translational alignment in their individual frames. Finally, the x-ray data here is in the low signal regime of Poissonian statistics, while the noise model for cryo-EM is not as simple.

The results in Fig~\ref{fig:results} suggest that there is a minimum number of photons/frame needed to determine the structure of the object. A similar feasibility criterion was found in the EMC simulations for single molecule diffraction imaging~\cite{loh026705}~\cite{elser2009}. In this case, the criterion depends on the particular object. More specifically, the more attenuation there is far from the rotation axis, the easier it is to assign orientations to frames. Also, a higher contrast object with the same shape would be easier to reconstruct.

The goal of reconstructing a single biomolecule or microcrystal needs a few more steps. First, we need to demonstrate a reconstruction with randomly-oriented diffraction data. Secondly, there would be sources of background we have not included, such as air-scatter, which would make the reconstruction more challenging.

\section{Acknowledgements}

Research on the development and application of x-ray detectors is supported by DOE Grant DE-FG02-10ER46693, the Keck Foundation, and CHESS. CHESS is supported by NSF and NIH-NIGMS under NSF Grant DMR-0936384. The data analysis work is supported by DOE Grant DE-FG02-11ER16210. The sample used to demonstrate tomographic reconstruction was graciously provided by Russell K. Philipp.

\end{document}